\documentclass[intlimits,twoside,a4paper]{article}

\usepackage{amsmath,amssymb}
\usepackage{graphicx}

\usepackage[T2A]{fontenc}
\usepackage[cp1251]{inputenc}

\usepackage{amsthm}
\usepackage{emlines}

\usepackage[eqsecnum]{cmpj2}

\newtheorem{remark}{Remark}

\issue{2013}{16}{2}{23701}

\doinumber{10.5488/CMP.16.23701}

\title[An exactly solvable quantum  superradiance model]%
{A new exactly solvable spatially one-dimensional quantum superradiance
fermi-medium model \\ and its  quantum solitonic states%
\thanks{Authors with great pleasure devote  their work to  Professor  Mykhaylo Kozlovskii in honor of his $60^{\mbox{th}}$ Birthday Jubilee.}}
\author[D. Blackmore, A. Prykarpatsky]{ D. Blackmore\refaddr{label1}, A. Prykarpatsky\refaddr{label2,label3} }
\addresses
{\addr{label1} New Jersey Institute of Technology, University Heights, Newark, NJ 07-102, USA
\addr{label2} AGH University of Science and Technology,  30-059 Krakow, Poland
\addr{label3} Ivan Franko State Pedagogical University, Lviv region, Drohobych, Ukraine}

\date{Received  October 23, 2012, in final form March 11, 2013}
\authorcopyright{D. Blackmore, A. Prykarpatsky, 2013}

\begin{document}

\maketitle

\begin{abstract}
A new exactly solvable spatially one-dimensional quantum superradiance model describing a charged fermionic
medium  interacting with an external electromagnetic field  is proposed.  The infinite hierarchy of quantum conservation laws
and many-particle  Bethe eigenstates that model quantum solitonic impulse  structures  are constructed. The    Hamilton  operator  renormalization
procedure subject to a physically stable vacuum is described, the   quantum excitations and quantum  solitons,  related to the thermodynamical   equilibrity of the model,   are  discussed.
\pacs 73.21.Fg, 73.63.Hs, 73.50.Fq, 78.67.De
\end{abstract}

\section{ Fermionic medium and superradiance model description}

We shall describe the quantum superradiance properties \cite%
{AE,PTB,CBKB,OUF,BCCDPP,GH,Fl,Mu,GOFPL} of a model of a one-dimensional many
particle charged fermionic medium interacting with an external
electromagnetic field. The Dirac type $N$-particle Hamiltonian operator of
the model is expressed as
\begin{equation}
H_{N}:=\ri\sum_{j=1}^{N}\sigma _{3}^{(j)}\frac{\partial }{\partial x_{j}}%
\otimes \mathbf{1-}\ri\beta \mathbf{1\otimes }\int_{\mathbb{R}}\rd x\varepsilon
^{+}\varepsilon _{x}+\alpha \sum_{j=1}^{N}\sigma _{1}^{(j)}\otimes \mathcal{E%
}(x_{j}),  \label{S1.1}
\end{equation}%
where $\sigma _{3}^{(j)},\sigma _{1}^{(j)},j=1,\ldots N,$ are the usual
Pauli matrices, $\alpha \in \mathbb{R}_{+}$ is an interaction constant, $%
0<\beta <1$ is the light speed in the linearly polarized fermionic medium,
\[%
\mathcal{E}(x):=\left(
\begin{array}{cc}
\varepsilon (x) & 0 \\
0 & \varepsilon ^{+}(x)%
\end{array}%
\right)
\]
is the one-mode polarization matrix operator at particle location $%
x\in \mathbb{R}$ with quantized electric field bose-operators $\varepsilon
(x),\varepsilon ^{+}(x):\Phi _\mathrm{B}\rightarrow \Phi _\mathrm{B}$ acting in the
corresponding Fock space $\Phi _\mathrm{B}$ and satisfying the commutation
relationships:%
\begin{equation}
\label{S1.2}
\lbrack \varepsilon (x),\varepsilon ^{+}(y)]=\delta (x-y),
\qquad \lbrack \varepsilon (x),\varepsilon (y)]=0=[\varepsilon ^{+}(x),\varepsilon
^{+}(y)]  \nonumber
\end{equation}%
for all $x,y\in \mathbb{R}.$ We note that throughout the sequel we employ
units for which the standard constants $\hslash =1=c.$

By construction, the $N$-particle Hamiltonian operator (\ref{S1.1}) acts
in the Hilbert space $L_{2}^{(\mathrm{as})}(\mathbb{R}^{N};\mathbb{C}^{2})\otimes
\Phi _\mathrm{B},$ where $L_{2}^{(\mathrm{as})}(\mathbb{R}^{N};\mathbb{C}^{2})$ denotes the
square-integrable antisymmetric vector functions on $\mathbb{R}^{N},N\in
\mathbb{Z}_{+}.$ Correspondingly, the Fock space $\Phi _\mathrm{B}$ allows for the
standard representation as the direct sum
\begin{equation}
\Phi _\mathrm{B}:=\bigoplus _{n\in \mathbb{Z}_{+}}L_{2}^{(\mathrm{s})}(\mathbb{R}^{n};\mathbb{C}%
),  \label{S1.3}
\end{equation}%
where $L_{2}^{(\mathrm{s})}(\mathbb{R}^{n};\mathbb{C})$ denotes the space of
symmetric square-integrable scalar functions on $\mathbb{R}^{n},n\in \mathbb{%
Z}_{+}.$ Similarly, the corresponding fermionic Fock space
\begin{equation}
\Phi _\mathrm{F}:=\bigoplus _{n\in \mathbb{Z}_{+}}L_{2}^{(\mathrm{as})}(\mathbb{R}^{n};\mathbb{C%
}^{2}),  \label{S1.4}
\end{equation}%
can be used to represent \cite{BB,Fe,Is,BPS} the Hamiltonian operator (\ref%
{S1.1}) in the second quantized form%
\begin{equation}
\mathbf{H:}=\ri\int_{\mathbb{R}}\rd x\left[\psi _{1}^{+}\psi _{1,x}-\psi _{2}^{+}\psi
_{2,x}-\beta \varepsilon ^{+}\varepsilon _{x}+\ri\alpha \left(\varepsilon \psi
_{2}^{+}\psi _{1}+\varepsilon ^{+}\psi _{1}^{+}\psi _{2}\right)\right],  \label{S1.5}
\end{equation}%
which acts on the tensored Fock space $\Phi :=\Phi _\mathrm{F}\otimes \Phi _\mathrm{B},$
where the spaces $\Phi _\mathrm{F}$  and $\Phi _\mathrm{B},$   defined respectively, by (\ref{S1.3}) and (\ref{S1.4}), can also be represented as
\begin{eqnarray}
\label{S1.6}
%
\Phi _\mathrm{B}&:=&\bigoplus _{n\in \mathbb{Z}_{+}}\mathrm{span}\Bigg\{ \int_{\mathbb{R}%
^{n}}\rd x_{1}\rd x_{2}\ldots \rd x_{n}\chi _{n}(x_{1},x_{2,}\ldots,x_{n})  \nonumber  \\
&& \times  \prod_{j=1}^{n}\varepsilon ^{+}(x_{j})\left.
|0\right\rangle _\mathrm{B};\chi _{n}\in L_{2}^{(\mathrm{s})}(\mathbb{R}^{n};\mathbb{C}%
)\Bigg\} ,%
\nonumber\\
\Phi _\mathrm{F} &:=&\bigoplus _{n\in \mathbb{Z}_{+}}\mathrm{span}\Bigg\{ \int_{%
\mathbb{R}^{n}}\rd x_{1}\rd x_{2}\ldots \rd x_{n}\varphi
_{n}^{(m)}(x_{1},x_{2,}\ldots,x_{n})\nonumber \\
&&\times \prod_{j=m+1}^{n}\psi _{1}^{+}(x_{j})\prod_{k=1}^{m}\left. \psi
_{2}^{+}(x_{k}) |0\right\rangle :0\leqslant m\leqslant n;\varphi _{n}^{(m)}\in
L_{2}^{(\mathrm{as})}(\mathbb{R}^{n};\mathbb{C}^{2})\Bigg\} ,
\end{eqnarray}
and $\left. |0\right\rangle _\mathrm{B}\in \Phi _\mathrm{B},\left. |0\right\rangle _\mathrm{F}\in
\Phi _\mathrm{F}$ are the corresponding vacuum bose- and fermi-states, satisfying
the determining conditions
\begin{equation}
\psi _{1}(x)\left. |0\right\rangle _\mathrm{F}=0=\psi _{2}(x)\left. |0\right\rangle
_\mathrm{F}, \qquad \varepsilon (x)\left. |0\right\rangle _\mathrm{B}=0  \label{S1.7}
\end{equation}%
for all $x\in \mathbb{R}.$ The creation and annihilation operators $\psi
_{j}(x),\psi _{k}^{+}(y):\Phi _\mathrm{F}\rightarrow \Phi _\mathrm{F},j,k=1,2,$  satisfy
the anti-commuting
\begin{align}
&\{\psi _{j}(x),\psi _{k}^{+}(y)\} =\delta _{j,k}\delta (x-y),  \label{S1.8a}
\nonumber
\\
&\{\psi _{j}(x),\psi _{k}(y)\}\, =0=\{\psi _{j}^{+}(x),\psi _{k}^{+}(y)\}
\end{align}%
and commuting
\begin{align}
&[ \varepsilon (x),\psi _{j}(y)] =0=[\varepsilon (x),\psi _{j}^{+}(y)],
\label{S1.8b} \nonumber\\
&[ \varepsilon ^{+}(x),\psi _{j}(y)] =0=[\varepsilon ^{+}(x),\psi
_{j}^{+}(y)]
\end{align}%
relationships for all $x,y\in \mathbb{R}$.

As we are interested in describing the so-called super-resonance processes in
our fermionic medium induced by an external electromagnetic field, in
particular,  a possibility of generating strong localized photonic impulses,
it is first necessary to  investigate the bound photonic medium states and
their spectral energy characteristics. Toward this aim, we make an important
note that it has been observed that the spectral properties of the
Hamiltonian operator (\ref{S1.5}) can be analyzed in great detail owing to
the fact that the related Heisenberg nonlinear dynamical system
\begin{align}
\psi _{1,t} & =\ri[\mathbf{H,}\psi _{1}]=\psi _{1,x}+\ri\alpha \varepsilon
^{+}\psi _{2},  \nonumber \\
\psi _{2,t}& =\ri[\mathbf{H,}\psi _{2}]=-\psi _{2,x}+\ri\alpha \varepsilon \psi
_{1},  \label{S1.9} \nonumber\\
\varepsilon _{t}& =\ri[\mathbf{H,}\varepsilon ]=-\beta \varepsilon
_{x}+\ri\alpha \psi _{1}^{+}\psi _{2}
\end{align}%
is a quantum exactly solvable Hamiltonian flow on the quantum operator
manifold $\mathbf{M}:=\{(\psi _{1},\psi _{2},\varepsilon ;\varepsilon
^{+},\psi _{2}^{+},\psi _{1}^{+})\in \mathrm{End}\Phi ^{6}\}$, where the
notation $\mathrm{End}\Phi ^{6}$ denotes the space of all endomorphisms from
the linear operator space $\Phi ^{6}$ to itself. The system (\ref{S1.9})
can be linearized by means of the quantum Lax type spectral problem%
\begin{equation}
\frac{\rd f}{\rd x}=l(x;\lambda )f,  \label{S1.10}
\end{equation}%
where the generalized eigenfunction $f\in \Phi ^{3},$ and the operator
matrix $l(x;\lambda )\in \mathrm{End}\Phi ^{3\text{ }}$equals%
\begin{equation}
l(x;\lambda ):=\left(
\begin{array}{ccc}
-\frac{\ri\lambda }{3-\beta } & \ri\xi _{1}\psi _{1} & \ri\xi _{2}\psi _{2} \\
\ri\xi _{1}\psi _{1}^{+} & -\frac{\ri\lambda }{2\beta } & \ri\xi _{3}\varepsilon
\\
\ri\xi _{2}\psi _{2}^{+} & \ri\xi _{3}\varepsilon ^{+} & \frac{\ri\lambda (3+\beta
)}{2\beta (3-\beta )}%
\end{array}%
\right)   \label{S1.11}
\end{equation}%
for all $x\in \mathbb{R},$ with $\lambda \in \mathbb{C}$ being an
arbitrary time-independent spectral parameter, and
\begin{align}
\xi _{1}& :=\xi _{1}(\alpha ,\beta )=-18\alpha \left[ \frac{(9-3\beta
)(\beta +1)}{\beta +3}\right] ^{1/2}
\left( \frac{12\beta }{\beta +3}\right) ^{1/2}\frac{\beta +3}{%
2\beta ^{2}+3\beta +3}\,,  \label{S1.12} \nonumber\\
\xi _{2}& :=\xi _{2}(\alpha ,\beta )=6\alpha (3-3\beta )^{1/2}\left( \frac{%
12\beta }{\beta +3}\right) ^{1/2}\frac{(9-3\beta )(\beta +1)}{(\beta
-1)(2\beta ^{2}+3\beta +3)}\,,  \nonumber \\
\xi _{3}& :=\xi _{3}(\alpha ,\beta )=72\alpha \beta \frac{(3(1-\beta ))^{1/2}%
}{(\beta -1)(2\beta ^{2}+3\beta +3)}\left[ \frac{(9-3\beta )(\beta +1)}{%
\beta +3}\right] ^{1/2}
\end{align}%
being  constants,  depending on the interaction parameter $\alpha \in
\mathbb{R}_{+}$ and the light speed $\beta $ in the polarized fermionic
medium $0<\beta <1.$

\begin{remark}
Concerning   the  quantum Lax type spectral problem  (\ref{S1.10}) and its
determination, one can consult  \cite{Fa,Ko,FT,BPS,MBPS}, where the corresponding
analytical tools are developed and described in detail.
\end{remark}

The quantum dynamical system (\ref{S1.5}) may be also regarded as an exactly
solvable approximation of the three-level quantum model studied in \cite{Bo}
subject to its superradiance properties. Concerning the studies of such
superradiance Dicke type one-dimensional models, it is necessary to mention
the work \cite{Ru} in which it was shown that the well-known quantum
Bloch-Maxwell dynamical system
\begin{align}
\psi _{1,t}& =\ri[\mathbf{\hat{H},}\psi _{1}]=\ri\alpha \varepsilon ^{+}\psi
_{2}\,,  \label{S1.13} \nonumber\\
\psi _{2,t}& =\ri[\mathbf{\hat{H},}\psi _{2}]=\ri\alpha \varepsilon \psi _{1}\,,
\nonumber \\
\varepsilon_{t}\phantom{_{2}}& =\ri[\mathbf{\hat{H},}\varepsilon ]=-\beta \varepsilon
_{x}+\ri\alpha \psi _{1}^{+}\psi _{2}
\end{align}%
generated by the the reduced quantum Hamiltonian operator
\begin{equation}
\mathbf{\hat{H}:}=-\ri\int_{\mathbb{R}}\rd x\left[\beta \varepsilon ^{+}\varepsilon
_{x}-\ri\alpha \left(\varepsilon \psi _{2}^{+}\psi _{1}+\varepsilon ^{+}\psi
_{1}^{+}\psi _{2}\right)\right]  \label{S1.14}
\end{equation}%
in the strongly degenerate Fock space $\Phi $ is also exactly solvable.
Moreover, it has the corresponding Lax type operator spectral problem \cite%
{FT,No,BPS} in the space $\Phi ^{3}.$ However, the important problem of
constructing the stable physical vacuum for the Hamiltonian (\ref{S1.14})
was on the whole not discussed in \cite{Ru}, and neither was the problem of
studying the related thermodynamics of quantum excitations over it. More
interesting quantum one-dimensional models with the Hamiltonian similar to (%
\ref{S1.5}) describing the quantum interaction of just fermionic particles
and only bosonic particles with an external electromagnetic field were
studied, respectively, in \cite{WO} and \cite{Ku}. In these investigations,
the quantum localized Bethe states were constructed and analyzed in detail.
The corresponding classical version of the quantum dynamical system (\ref%
{S1.9}), called the \textit{three-wave model}, was studied in \cite{No,ZM,Oh}
and elsewhere.

It is also worthy to  mention  here that the spectral operator problem (\ref%
{S1.11}) makes sense  \cite{GH,GOFPL,OUF} only if the light speed inside the
polarized fermionic medium is less than the light speed in a vacuum. This is
ensured by the dynamical stability of the quantum Hamiltonian system  (%
\ref{S1.9}) following from  the existence of an additional infinite
hierarchy of conservation laws, suitably determined on the quantum operator
phase space $\mathbf{M}.$ Consequently, one can expect that the quantum
dynamical system (\ref{S1.9}) also possesses  many-particle localized
photonic states in the Fock space $\Phi ,$ which are called  \cite%
{Ru,PTB,Ru1,Ru2,Ru3,Ru4,Ru5} \textit{quantum solitons}, whose spatial range
is inverse to the number of interior particles, and which can be interpreted
as special Dicke type superradiance laser impulses. In particular, the
quantum stability, solitonic formation aspects and construction of the
physical ground state  \cite{BPS,MBPS,Fa} related to the unbounded {%
a priori} from below Hamiltonian operator (\ref{S1.5}) will be the main
focus of the succeeding sections.

\section{Bethe eigenstates and the energy localization}

In this section we construct some of the finite-particle Bethe eigenstates \cite{Fa,Ga,BPS,Ko,MBPS} for the quantum Hamiltonian operator  (\ref{S1.5}) and discuss their energy
localization property. The localization manifests itself in the fact that
the energy of a many-particle cluster appears to be less \cite{Fa,Ga,BPS} than that
of the corresponding system of free particles, giving rise to the formation
of the so-called quantum solitonic states localized in the space.

The following number operators
\begin{equation}
\mathbf{N}_\mathrm{F}:=\int_{\mathbb{R}}\rd x\left(\psi_{1}^{+}\psi_{1}+\psi_{2}^{+}\psi
_{2}\right),\mathbf{N}_\mathrm{B}:=\int_{\mathbb{R}}\rd x\left(\varepsilon^{+}\varepsilon+\psi
_{2}^{+}\psi_{2}\right)   \label{S2.1}
\end{equation}
commute with each other and with the Hamiltonian operator (\ref{S1.5}):%
\begin{equation}
\lbrack\mathbf{N}_\mathrm{F},\mathbf{N}_\mathrm{B}]=0, \qquad [\mathbf{H,N}_\mathrm{F}]=0=[\mathbf{H,N}%
_\mathrm{B}].   \label{S2.2}
\end{equation}
Hence, the Bethe eigenstates of the Hamiltonian operator  (\ref{S1.5}) can
be indexed by two integers $N,M\in\mathbb{Z}_{+}:$ the state $\left.
|(N,M)\right\rangle \in\Phi$ satisfies the determining equation
\begin{equation}
\mathbf{H}\left. |(N,M)\right\rangle =E\left. |(N,M)\right\rangle ,
\label{S2.3}
\end{equation}
where $E\in\mathbb{R}$ is the energy and
\begin{align}
\mathbf{N}_\mathrm{F}\left. |(N,M)\right\rangle & =N\left. |(N,M)\right\rangle ,
\label{S2.4} \nonumber\\
\mathbf{N}_\mathrm{B}\left. |(N,M)\right\rangle & =M\left. |(N,M)\right\rangle .
\end{align}
Owing to  (\ref{S2.4}), the state%
\begin{eqnarray}
\left. |(N,M)\right\rangle &=&\bigoplus_{n=0}^{N}\int_{\mathbb{R}%
^{N+M}}\rd x_{1}\rd x_{2}\ldots \rd x_{n}\rd y_{1}\rd y_{2}\ldots \rd y_{N-n}\rd z_{1}\rd z_{2}\ldots \rd z_{M-N+n} \nonumber\\
&&\times%
\varphi_{(M,N)}^{(n)}(x_{1},x_{2},\ldots,x_{n};y_{1},y_{2},y_{3},\ldots,y_{N-n};z_{1},z_{2},\ldots,z_{M-N+n})\nonumber\\
&&\times\prod_{j=1}^{n}\psi_{1}^{+}(x_{j})\prod_{k=1}^{N-n}\psi_{2}^{+}(y_{k})%
\prod_{l=1}^{M-N+n}\varepsilon^{+}(z_{l})\left. |0\right\rangle %
\label{S2.5}
\end{eqnarray}
for any $M,N\in\mathbb{Z}_{+}$, where for bounded states $\varphi
_{(M,N)}^{(n)}\in L_{2}^{(\mathrm{as})}(\mathbb{R}^{N};\mathbb{C})\times L_{2}^{(\mathrm{s})}(%
\mathbb{R}^{M-N+n};\mathbb{C}),0\leqslant n\leqslant N,$ one can easily construct \cite%
{Ga,Ko,WO} the Bethe state%
\begin{equation}
\left. |(1,1)\right\rangle =\int_{\mathbb{R}}\rd y_{1}\varphi(y_{1})%
\psi_{2}^{+}(y_{1})\left. |0\right\rangle +\int_{\mathbb{R}}\rd x_{1}\int_{%
\mathbb{R}}\rd z_{1}\chi(x_{1};z_{1})\psi_{1}^{+}(x_{1})\varepsilon^{+}(z_{1})%
\left. |0\right\rangle .   \label{S2.6}
\end{equation}
Here, the functions $\varphi:\mathbb{R\rightarrow C}$ and $\chi:\mathbb{R}%
^{2}\rightarrow\mathbb{C}$ satisfy the generalized differential equations%
\begin{eqnarray}
&&\phantom{-}\frac{1}{\ri}\frac{\partial\varphi(x_{1})}{\partial x_{1}}-\alpha\chi
(x_{1};x_{1}) =E\varphi(x_{1}),  \label{S2.7} \nonumber\\
&&-\frac{1}{\ri}\frac{\partial\chi(x_{1};x_{2})}{\partial x_{1}}+\frac{\beta}{\ri}%
\frac{\partial\chi(x_{1};x_{2})}{\partial x_{2}}-\alpha\varphi(x_{1})%
\delta(x_{1}-x_{2}) =E\chi(x_{1};x_{2}),
\end{eqnarray}
having solutions
\begin{align}
\varphi(x_{1}) & =S_{+}(\lambda,\mu)\exp\left[\ri\left(p_{1}+q_{1}\right)x\right],  \label{S2.8} \nonumber \\
\chi(x_{1},x_{2}) & =\left[\vartheta\left(x_{1}-x_{2}\right)+S_{1}(\lambda,\mu
)\vartheta(x_{2}-x_{1})\right]\exp\left[\ri\left(p_{1}x_{1}+q_{1}x_{2}\right)\right],
\end{align}
where the momenta $\lambda,\mu\in\mathbb{R}$ and
\begin{align}
p_{1} &=(\beta-1)\lambda, \qquad\qquad q_{1}=2\mu,\qquad \qquad E=p_{1}-\beta q_{1},  \label{S2.9} \nonumber\\
S_{1}(\lambda,\mu) &=\frac{\lambda-\mu-\ri{\alpha^{2}}/[4(1-\beta^{2})]}{%
\lambda-\mu+\ri{\alpha^{2}}/[4(1-\beta^{2})]}\,,\qquad S_{+}(\lambda,\mu )=\frac{%
\alpha}{2(1-\beta)\lambda-\mu+\ri{\alpha^{2}}/[4(1-\beta^{2})]}\,.
\end{align}
In the same way, one can represent the other quantum Bethe state as
\begin{equation}
\left. |(2,1)\right\rangle =\int_{\mathbb{R}^{2}}\rd x_{1}\rd x_{2}\varphi
(x_{1},x_{2})\psi_{1}^{+}(x_{1})\psi_{2}^{+}(x_{2})\left. |0\right\rangle
+\int_{\mathbb{R}^{3}}\rd x_{1}\rd x_{2}\rd x_{3}\chi(x_{1},x_{2};x_{3})%
\psi_{1}^{+}(x_{1})\varepsilon^{+}(x_{2})\left. |0\right\rangle ,
\label{S2.10}
\end{equation}
where the functions $\varphi:\mathbb{R}^{2}\mathbb{\rightarrow C}$ and $\chi:%
\mathbb{R}^{3}\rightarrow\mathbb{C}$ satisfy the generalized differential
equations:%
\begin{eqnarray}
&&-\frac{1}{\ri}\left[ \frac{\partial\varphi(x_{1},x_{2})}{\partial x_{1}}-\frac{%
\partial\varphi(x_{1},x_{2})}{\partial x_{2}}\right] +\alpha\chi
(x_{2},x_{1};x_{2})+\alpha\chi(x_{1},x_{2};x_{1})=E\varphi(x_{1},x_{2}), \nonumber\\
&&-\frac{1}{\ri}\left[ \frac{\partial\chi(x_{1},x_{2};x_{3})}{\partial x_{1}}+%
\frac{\partial\chi(x_{1},x_{2};x_{3})}{\partial x_{2}}\right] +\frac{\beta }{%
\ri}\frac{\partial\chi(x_{1},x_{2};x_{3})}{\partial x_{3}} \nonumber\\
&&+\frac{\alpha}{2}\varphi(x_{1},x_{2})\delta(x_{2}-x_{3})+\frac{\alpha}{2}%
\varphi(x_{2},x_{1})\delta(x_{1}-x_{3})=E\chi(x_{1},x_{2};x_{1}),%
\label{S2.11}
\end{eqnarray}
with solutions similar to those of (\ref{S2.8}) and (\ref{S2.9}).

It is important to mention here that the eigenstates (\ref{S2.6}) and (\ref%
{S2.10}) become degenerate as $\beta \rightarrow 1,$ meaning that the
corresponding bound quantum soliton states cannot be formed. The same
statement is also true for an arbitrary eigenstate (\ref{S2.5}). To
demonstrate this, we shall in the next section make use of the quantum
spectral problem (\ref{S1.11}) to prove that the quantum dynamical system (%
\ref{S1.9}) possesses an infinite hierarchy of commuting conservation laws,
thereby ensuring its complete quantum integrability and the formation
of quantum solitons.

\section{The quantum solitons}

We now consider the following quantum operator Cauchy problem for the
spectral equation (\ref{S1.10}) subject to the periodic conditions $l(x+2\pi
;\lambda)=l(x;\lambda)\in\mathrm{End}\Phi^{3}$ for all $x\in$ $\mathbb{R}$
and $\lambda\in\mathbb{C}:$%
\begin{equation}
\frac{\rd F(x,y;\lambda)}{\rd x}=\vdots l(x;\lambda)F(x,y;\lambda)\vdots\,,   \label{S3}
\end{equation}
where $F(x,y;\lambda)\in\mathrm{End}\Phi^{3}$ is the corresponding
fundamental transition operator matrix satisfying
\begin{equation}
\left. F(x,y;\lambda)\right\vert _{y=x}=1,   \label{S3.2}
\end{equation}
and the operation $\vdots\cdot\vdots$ arranges operators $\psi_{j,}\psi
_{j}^{+},j=1,2,$ $\varepsilon$ and $\varepsilon^{+},$ via the standard
normal ordering \cite{Sk,BB} that does not change the position of any other
operators; for instance, $\vdots A\psi_{1}^{+}\psi_{2}\varepsilon^{+}B\vdots
$ $=\psi_{1}^{+}\varepsilon^{+}AB\psi_{2}$ for any $A,B\in\mathrm{End}\Phi.$

Construct now the operator products
\begin{eqnarray}
\mathcal{\hat{F}}(x,y|\lambda,\mu):=\tilde{F}(x,y;\lambda)\overset{%
\thickapprox}{F}(x,y;\mu),  \nonumber \\\label{S3.3}
\mathcal{\check{F}}(x,y|\lambda,\mu):=\overset{\thickapprox}{F}(x,y;\mu )%
\tilde{F}(x,y;\lambda),
\end{eqnarray}%
where
\begin{eqnarray}
\tilde{F}(x,y;\lambda) & :=F(x,y;\lambda)\otimes1,  \label{S3.4} \nonumber \\
\overset{\thickapprox}{F}(x,y;\mu) & :=1\otimes F(x,y;\mu)
\end{eqnarray}
are for all $x,y\in\mathbb{R}$ $,$ $\lambda,\mu\in\mathbb{C},$ the
corresponding tensor products of operators acting in the space $\Phi
^{3}\otimes\Phi^{3}.$ The following proposition is crucial \cite{Sk,BPS,MBPS}
for further analysis of integrability of the quantum
dynamical system (\ref{S1.9}) and is proved by a direct computation.

\medskip

\textbf{Proposition 1.} \textit{The operator expressions (\ref%
{S3.3}) satisfy the following differential relationships: }%
\begin{align}
\frac{\partial }{\partial x}\mathcal{\hat{F}}(x,y|\lambda ,\mu )& =\vdots
\mathcal{\hat{L}}(x;\lambda ,\mu )\mathcal{\hat{F}}(x,y|\lambda ,\mu )\vdots
\,,  \label{S3.5} \nonumber\\
\frac{\partial }{\partial x}\mathcal{\check{F}}(x,y|\lambda ,\mu )& =\vdots
\overset{\smallsmile }{\mathcal{L}}\mathcal{(}x;\lambda ,\mu )\mathcal{%
\check{F}}(x,y|\lambda ,\mu )\vdots \,,
\end{align}%
where the matrices%
\begin{align}
\mathcal{\hat{L}(}x;\lambda ,\mu )& =\tilde{l}(\text{ }x;\lambda )+\overset{%
\approx }{l}(x;\mu )-\alpha \hat{\bigtriangleup}(x;\lambda ,\mu )\,,
\label{S3.6} \nonumber\\
\overset{\smallsmile }{\mathcal{L}}\mathcal{(}x;\lambda ,\mu )& =\tilde{l}(%
\text{ }x;\lambda )+\overset{\approx }{l}(x;\mu )-\alpha \check{%
\bigtriangleup}(x;\lambda ,\mu )\,,
\end{align}%
and $\hat{\bigtriangleup}(x;\lambda ,\mu ),\check{\bigtriangleup}(x;\lambda
,\mu )$ satisfy the algebraic relationship $P\hat{\bigtriangleup}(x;\lambda
,\mu )P=\check{\bigtriangleup}(x;\lambda ,\mu )$  for all $x\in R,$ $%
\lambda ,\mu \in C,$ where $P\in \mathrm{End} \Phi ^{3}\otimes \Phi ^{3}$\ is the
standard transmutation operator in the space $\Phi ^{3}\otimes \Phi ^{3},$
that is $P(a\otimes b):=b\otimes a$ for any vectors $a,b\in \Phi ^{3}.$

Using proposition~1, one can easily verify that there exists a scalar $%
\mathcal{R}$-matrix $\mathcal{R(\lambda },\mathcal{\mu )},$ $\mathcal{R} \in%
\mathrm{End}\mathbb{C}^{9},$ such that
\begin{equation}
\mathcal{R(\lambda },\mu )\mathcal{\hat{L}(}x;\lambda ,\mu )=\overset{%
\smallsmile }{\mathcal{L}}\mathcal{(}x;\lambda ,\mu )\mathcal{R(\lambda }%
,\mu )  \label{S3.7}
\end{equation}%
holds for all $\lambda ,\mu \in \mathbb{C}$ and $x\in \mathbb{R}.$ This,
owing to the equations (\ref{S3.5}), implies the main functional
Yang-Baxter type \cite{Fa,Sk,MBPS} operator relationship%
\begin{equation}
\mathcal{R(\lambda },\mu )\mathcal{\hat{F}}(x,y|\lambda ,\mu )=\mathcal{%
\check{F}}(x,y|\lambda ,\mu )\mathcal{R(\lambda },\mu )  \label{S3.8}
\end{equation}%
is satisfied for any $x,y\in \mathbb{R}$ and $\lambda ,\mu \in \mathbb{C},$
where, as a result of \cite{Ku},
\begin{equation}
\mathcal{R(\lambda },\mu )=(\lambda -\mu )P-\ri\alpha \mathbf{1.}  \label{S3.9}
\end{equation}%
Recalling now the periodicity condition, from (\ref{S3.8}), one easily
deduces by means of the trace-operation that the monodromy operator matrix $%
T(x;\lambda ):=F(x+2\pi ,x;\lambda )$ satisfies the following commutation
relationship for all $x\in $ $\mathbb{R}$ and $\lambda ,\mu \in $ $\mathbb{C}%
:$%
\begin{equation}
\left[ \mathrm{tr}T(x;\lambda ),\mathrm{tr}T(x;\mu )\right] =0.
\label{S3.11}
\end{equation}%
Actually, it follows from (\ref{S3.8}) that
\begin{align}
\mathrm{tr}\left\{T(x;\lambda )\otimes T(x;\mu )\right\}& =\mathrm{tr}\left\{\mathcal{R}%
^{-1}T(x;\mu )\otimes T(x;\lambda )\right\}=\mathrm{tr}\left\{T(x;\mu )\otimes T(x;\lambda )\right\}.  \label{S3.12}
\end{align}%
Taking into account that $\mathrm{tr}(A\otimes B)=\mathrm{tr}A\cdot \mathrm{%
tr}B$ for any operators $A,B\in \mathrm{End} \Phi ^{3},$ one easily obtains
(\ref{S3.11}) from (\ref{S3.12}). Consequently, the $\lambda$-dependent
operator functional
\begin{equation}
\gamma (\lambda ):=\mathrm{tr}T(x,\lambda )\cong \Sigma _{j\in \mathbb{Z}%
_{+}}\gamma _{j}\lambda ^{-j},  \label{S3.13}
\end{equation}%
as $\left\vert \lambda \right\vert \rightarrow \infty $ generates an
infinite hierarchy of commuting conservation laws $\gamma _{j}:\Phi
\rightarrow \Phi ,j\in \mathbb{Z}_{+}$ $:$%
\begin{equation}
\left[ \gamma _{j},\gamma _{k}\right] =0  \label{S3.14}
\end{equation}%
for all $j,k\in \mathbb{Z}_{+},$ where, in particular,
\begin{align}
\gamma _{1}& =\mathbf{N}_\mathrm{F}=\int_{\mathbb{R}}\rd x(\psi _{1}^{+}\psi _{1}+\psi
_{2}^{+}\psi _{2}), \qquad \gamma _{2}=\mathbf{N}_\mathrm{B}=\int_{\mathbb{R}%
}\rd x(\varepsilon ^{+}\varepsilon +\psi _{2}^{+}\psi _{2}),  \label{S3.15} \nonumber\\
\gamma _{3}& =\mathbf{P}=\ri\int_{\mathbb{R}}\rd x(\psi _{1}^{+}\psi _{1,x}+\psi
_{2}^{+}\psi _{2,x}+\varepsilon ^{+}\varepsilon _{x}),  \nonumber \nonumber\\
\gamma _{4}& =\mathbf{H}=\ri\int_{\mathbb{R}}\rd x\left[ \psi _{1}^{+}\psi
_{1,x}-\psi _{2}^{+}\psi _{2,x}-\varepsilon ^{+}\varepsilon _{x}+\ri\alpha
(\varepsilon \psi _{2}^{+}\psi _{1}+\psi _{1}^{+}\psi _{2}\varepsilon ^{+})%
\right] .
\end{align}%
Since the operator functional $\gamma _{4}=\mathbf{H}$ is the Hamiltonian
operator for the dynamical system (\ref{S1.9}), from (\ref{S3.14}) one
obtains
\begin{equation}
\left[ \mathbf{H},\gamma _{j}\right] =0  \label{S3.16}
\end{equation}%
for all $j\in $ $\mathbb{Z}_{+}$; that is, all of functionals $\gamma
_{j}:\Phi \rightarrow \Phi ,j\in \mathbb{Z}_{+},$ are conservation laws.

Moreover, making use of the exact operator relationships (\ref{S3.8}) one
can construct the physically stable quantum states $|(N,M)\rangle\in\Phi$ for all $%
N,M\in\mathbb{Z}_{+}$ upon redefining the Fock vacuum $\left.
|0\right\rangle \in\Phi,$ which is nonphysical for the dynamical system (\ref%
{S1.9}), governed by the unbounded from below Hamiltonian operator (\ref%
{S1.5}). Following a renormalization scheme similar to those developed in
\cite{Fa,YK,MBPS}, one can construct a new physically stable vacuum
\begin{equation}
\left. |0\right\rangle _\mathrm{phys}:=\prod_{q\leqslant\mu_{j}\leqslant Q}B^{+}(\mu_{j})\left.
|0\right\rangle   \label{S3.17}
\end{equation}
by means of the new, commuting to each other, ``creation'' operators $%
B^{+}(\mu):\Phi\rightarrow\Phi ,\mu\in\mathbb{C},$ generated by suitable
components of the monodromy operator matrix $T(x;\mu):\Phi^{3}\rightarrow%
\Phi^{3},$ $x\in\mathbb{R},$ whose commutation relationships with the
Hamiltonian operator (\ref{S1.5})
\begin{equation}
\lbrack\mathbf{H,}B^{+}(\mu)]=S(\mu;\alpha,\beta)B^{+}(\mu)   \label{S3.18}
\end{equation}
are parameterized by the two-particle scalar scattering factor $S(\mu
;\alpha,\beta),\mu\in\mathbb{C},$ and where the values $q<Q\in\mathbb{R}$ are to
be determined from the condition that quantum excitations over the physical
vacuum  (\ref{S3.17}) have positive energy. Since the physical vacuum (\ref%
{S3.17}) is an eigenstate of the Hamiltonian operator (\ref{S1.5}), the
corresponding quantum eigenstates of the excitations can be represented as
\begin{equation}
\left. |(\mu)\right\rangle :=B^{+}(\mu)\left. |0\right\rangle _\mathrm{phys}
\label{S3.19}
\end{equation}
for some $\mu\in\mathbb{R}$ and the new energy level can be taken into
account in the renormalized Hamiltonian operator (\ref{S1.5}) by means of
the chemical potentials $a_\mathrm{F},a_\mathrm{B}\in\mathbb{R}:$%
\begin{equation}
\mathbf{H}_{a}:=\mathbf{H}-a_\mathrm{F}\mathbf{N}_\mathrm{F}-a_\mathrm{B}\mathbf{N}_\mathrm{B}\,,
\label{S3.20}
\end{equation}
which should be determined from the conditions
\begin{equation}
\mathbf{H}_{a}\left.|0\right\rangle_\mathrm{phys} =0,\qquad\left\langle (\mu)\right. |%
\mathbf{H}_{a}\left.|(\mu)\right\rangle >0   \label{S3.21}
\end{equation}
for any $\mu\in\mathbb{R}.$ The physical vacuum state and quantum
Hamiltonian renormalization construction described above make it possible to
study the properties of superradiance quantum photonic impulse structures
generated by interaction of the charged fermionic medium with an external
electromagnetic field. Owing to the existence of quantum periodic
eigenstates over the physically stable vacuum, one can also investigate the
related thermodynamic properties of the model and analyze the generated
superradiance photonic structures, which are important for explaining many
\cite{AE} existing experiments.

\section{Conclusion}

The exact solvability of our model describing a one-dimensional many-particle charged fermionic medium interacting with an external
electromagnetic field allows one to calculate diverse superradiance effects,
which are closely related to the formation of the bound quantum solitonic
states and their stability. The existence of the bound states is established by
suitably applying the physical vacuum renormalization subject to which all
quantum excitations are of positive energy. This procedure, based on the
determining operator relationships (\ref{S3.8}), (\ref{S3.17}) and (\ref{S3.18}) enables one to
describe the thermodynamic properties of the quantum dynamical system over
a stable physical vacuum. In addition, it facilitates the analysis of the
corresponding thermodynamic states of the resulting quantum photonic system
and its superradiance properties. Our results indicate  that a more detailed
investigation of these and related topics is in order, which we plan to
undertake elsewhere.

\section*{Acknowledgements}

A.P. is cordially thankful to Prof. J.~Slawianowski (IPPT of PAN, Warsaw)
for his invitation to deliver a report for his Seminar at IPPT of PAN, his
gracious hospitality, fruitful discussions and valuable remarks. D.B. was partially supported by NSF Grant CMMI--1029809. Last but
not least, cordial thanks belong to the referees, whose professional reading
of the original version of the manuscript was instrumental in making the
exposition of the results both clearer, correct and readable.

\ukrainianpart

\title{Нова точно розв'язувана просторово-одновимірна квантова модель супервипромінювального  фермі-середовища та квантові солітонні стани}
\author{Д. Блекмор\refaddr{label1}, А.К. Прикарпатський\refaddr{label2,label3}}
\addresses{
\addr{label1} Інститут науки та технологій, Університетські висоти, Ньюарк, 07-102 Нью Джерсі,  США
\addr{label2} Академія гірництва та металургії, 30-59 Краків,  Польща
\addr{label3} Державний педагогічний університет ім. І.~Франка, Дрогобич, Україна
}

\makeukrtitle

\begin{abstract}
\tolerance=3000%
Запропоновано нову точно розв'язувану просторово-одновимірну квантову супервипромінювальну  модель, що описує заряджене ферміонне середовище,  взаємодіюче із зовнішнім електромагнітним полем.  Сконструйовано скінченну ієрархію законів збереження та багаточастинкові власні стани Бете, що моделюють квантові солітонні імпульсні структури. Описано  процедуру ренормалізації оператора Гамільтона щодо стійкого фізичного вакуума, обговорюються  квантові збудження та  квантові солітони,    асоційовані  із термодинамічною рівновагою  моделі.
\keywords    заряджене ферміонне середовище,     квантова обернена  спектральна проблема, власні стани Бете,   ренормалізація вакуума, квантова супервипрмінювальна модель, закони збереження, тотожність Янга-Бакстера

\end{abstract}

\end{document}